\documentclass[runningheads]{llncs}

\usepackage[utf8]{inputenc}
\usepackage[T1]{fontenc}
\usepackage{lmodern}
\usepackage{multirow}
\usepackage{lscape}
\usepackage{amsmath}

\usepackage{enumerate}
\usepackage{amsfonts}
\usepackage{amssymb}
\usepackage{graphicx}
\usepackage{mathrsfs}
\usepackage{float}
\usepackage[]{algorithmic}
\usepackage{algorithm}
\usepackage[USenglish,british,american,australian,english]{babel}
\usepackage{setspace}
\usepackage[textfont={it}]{caption}
\usepackage{cases}
\usepackage{hyperref}
\usepackage[svgnames,dvipsnames]{xcolor}

\newcommand{\keywords}[1]{\par\addvspace\baselineskip
\noindent\keywordname\enspace\ignorespaces#1}
\newenvironment{example1}{\noindent\textbf{Example 1 \ }\itshape}{\hfill$\Box$\medskip\par }

\newcommand{\bb}[1]{\mathbb{#1}}
\renewcommand{\cal}[1]{\mathcal{#1}}

\newenvironment{preuve}{\noindent\textit{Proof. \ }}{\hfill$\Box$\medskip\par }
\newtheorem{lemme}{Lemma}
\newcommand{\pqcr}{\texttt{PQCR}}
\newcommand{\bab}{branch-and-bound}
\title{Solving unconstrained 0-1 polynomial programs through quadratic convex reformulation}

\date{\today}

\author{Sourour Elloumi\inst{1,2}, Am\'elie Lambert\inst{2} and Arnaud Lazare\inst{1,2}}

\authorrunning{Sourour Elloumi,  Am\'elie Lambert and Arnaud Lazare}

\institute{$1$ UMA-ENSTA,  828 Boulevard des Mar\'echaux, 91120 Palaiseau, France\\
\email{ \{sourour.elloumi,arnaud.lazare\}@ensta-paristech.fr}\\
$2$ CEDRIC-Cnam, 292 rue saint Martin, F-75141 Paris Cedex 03, France\\
\email{amelie.lambert@cnam.fr}
}

\toctitle{Solving unconstrained 0-1 polynomial programs}
\titlerunning{Solving unconstrained 0-1 polynomial programs}

\begin{document}

\maketitle

\paragraph{Abstract}
We propose an exact solution approach for the problem $(P)$ of minimizing an unconstrained binary polynomial optimization problem. We call \pqcr\\ (\texttt{Polynomial Quadratic Convex Reformulation}) this three-phase method. The first phase consists in reformulating $(P)$ into a quadratic program $(QP)$. To that end, we recursively reduce the degree of $(P)$ to two, by use of the standard substitution of the product of two variables by a new one. We then obtain a linearly constrained binary quadratic program. In the second phase, we rewrite the objective function of $(QP)$ into an equivalent and parameterized quadratic function using the identity $x_i^2=x_i$ and other valid quadratic equalities that we introduce from the reformulation of phase 1. Then, we focus on finding the best parameters to get a quadratic convex program which continuous relaxation's optimal value is maximized. For this, we build a new semi-definite relaxation $(SDP)$ of $(QP)$. Then, we prove that the standard linearization inequalities, used for the quadratization step, are redundant in presence of the new quadratic equalities. Next, we deduce our optimal parameters from the dual optimal solution of $(SDP)$. The third phase consists in solving $(QP^*)$, the optimally reformulated problem, with a standard solver. In particular, at each node of the \bab, the solver computes the optimal value of a continuous quadratic convex program. We present computational results where we compare \pqcr~ with other convexification methods, and with the solver \texttt{Baron} \cite{baron}. We evaluate our method on instances of the image restoration problem \cite{CraRo17} and the low auto-correlation binary sequence problem \cite{Berna86} from \texttt{minlplib} \cite{minlplib}. For this last problem, 33 instances among the 45 were unsolved in \texttt{minlplib}. We solve to optimality 6 of them, and for the 27 others we improve primal and/or dual bounds.


\section{Introduction}

In this paper, we are interested in solving the unconstrained binary polynomial optimization problem that can be stated as follows:

\begin{numcases}{(P)}
\min  f(x)=\sum_{p=1}^mc_p\prod_{i\in\cal{M}_p}x_i\nonumber\\
\mbox{s.t.}  \nonumber\\
 \quad  x_i \in \{0,1\}, \quad i\in I    \nonumber
 \nonumber
\end{numcases} 

\noindent where $I=\{1,..,n\}$, $f(x)$ is an $n-$variable polynomial of degree $d$ and $m$ is the number of monomials. For a monomial $p$, $\mathcal{M}_p$ is the subset of $I$ containing the indexes of the variables involved in $p$. It follows that $d=\max_p\vert \cal{M}_p\vert$.

Unconstrained binary polynomial optimization is a general model that allows to formulate many important problems in optimization. The special case where the polynomial objective function of $(P)$ is a quadratic function (i.e. $d=2$) has been widely studied. In this case, $(P)$ has many applications, including those from financial analysis~\cite{Lau70}, cluster analysis~\cite{Rao71}, computer aided design~\cite{KraPru78} or machine scheduling~\cite{Rhy70}. Moreover, many graph combinatorial optimization problems such as determining maximum cliques, maximum cuts, maximum vertex packing or maximum independent sets can be formulated as quadratic optimization problems~\cite{BalPro11,BorHamSun91,ParXue94}. In the cubic case (i.e. $d=3$), the important class of satisfiability problems known as 3-SAT, can be formulated as $(P)$~\cite{Karp1972}. In the case where $d\geq 3$, there also exists many applications including, for example: the construction of binary sequences with low aperiodic correlation~\cite{Berna86} that is one of the most challenging problems in signal design theory, or the image restoration problem in computer vision~\cite{CraRo17}. 

Problem $(P)$ is $\cal{NP}$-hard~\cite{GarJoh79}. Practical difficulties come from the non-convexity of $f$ and the integrality of its variables. During the last decade, several algorithms that can handle $(P)$ were introduced. In particular, methods were designed to solve the more general class of mixed-integer nonlinear programs. These methods are \bab~algorithms based on a convex relaxation of $(P)$. More precisely, in a first step a convex relaxation is designed and then a \bab~is performed based on this relaxation. The most classical relaxation consists in the complete linearization of $(P)$, but quadratic convex relaxations can also be used. For instance, the well known $\alpha-$\bab~\cite{ADFN98} computes convex under-estimators of nonlinear functions by perturbing the diagonal of the Hessian matrix of the objective function. Several implementations of these algorithms are available, see for instance \texttt{Baron}~\cite{baron}, \texttt{Antigone}~\cite{antigone}, \texttt{SCIP}~\cite{scip} or \texttt{Couenne}~\cite{BLL09}.

In the case where the objective function is a polynomial, but the variables are continuous,  Lasserre proposes in~\cite{LASSERRE15} an algorithm based on a hierarchy of semi-definite relaxations of $(P)$. The idea is, at each rank of the hierarchy, to successively tighten semi-definite relaxations of $(P)$ in order to reach its optimal solution value. It is also proven in~\cite{LASSERRE15} that this hierarchy converges in a finite number of iterations to the optimal solution of the considered problem. Further, this work has been extended to hierarchies of second order conic programs~\cite{AhmMaj14,Ghaddar2016,Ghaddar2015}, and of sparse doubly non-negative relaxation~\cite{BBCPOP}. Although these algorithms were not originally tailored for binary programming, they can handle $(P)$ by considering the quadratic constraint $x_i^2=x_i$. Methods devoted to the binary polynomial case were also proposed. In~\cite{BucDam13,LasTha13}, the authors use separable or convex under-estimators to approximate a given polynomial. Other methods based on linear reformulations can be found in~\cite{CraRo17,For59,SheTun92}, in which linear equivalent formulations to $(P)$ are proposed and then improved. In \cite{BucRin07}, the authors focus on a polyhedral description of the linearization of a binary polynomial program. Finally, the work in~\cite{AnBoCrGr17} considers \textit{quadratizations} with a minimal number of additional variables.

In this paper, we focus on finding equivalent quadratic convex formulations of $(P)$. Quadratic convex reformulation methods~\cite{BilEll07,BEL16} were introduced for the specific case where $d=2$. The idea of these approaches is to build tight equivalent reformulations to $(P)$ that have a convex objective function. This equivalent problem can be built using the dual solution of a semi-definite relaxation of $(P)$, and further solved by a \bab~algorithm based on quadratic convex relaxation. Here, we consider the more general case where $d\geq 3$, and we propose to compute an equivalent convex formulation to $(P)$.  Hence, we present an exact solution method for problem $(P)$ that can be split in three phases. The first phase consists in building an equivalent formulation to $(P)$ where both objective function and constraints are at most quadratic. For this, we need to add some auxiliary variables. We then obtain problem $(QP)$ that has a quadratic objective function and linear inequalities.

Then in the second phase, we focus on the convexification of the obtained problem. As illustrated in the experiments of Section~4, the original \texttt{QCR} and \texttt{MIQCR} methods are not able to handle $(QP)$. Indeed, \texttt{QCR} leads to a reformulation with a weak bound, and in method  \texttt{MIQCR} the semi-definite program that we need to solve is too large. This is why, in this paper we introduce a tailored convexification phase. The idea is to apply convex quadratic reformulation to any quadratization of $(P)$. For this, we need null quadratic functions on the domain of $(QP)$ so as to perturb the Hessian matrix of the new quadratic objective function. One of these null functions comes from the classic binary identity, $x_i^2=x_i$. One contribution of this paper is the introduction of new null quadratic functions on the domain of $(QP)$. This set of functions varies according to the quadratization used in phase 1. Adding these functions to the new objective function, we get a family of convex equivalent formulations to $(QP)$ that depend on some parameters. We then want to choose these parameters such that the continuous relaxation bound of the convexified problem is maximized. We show that they can be computed thanks to a semi-definite program. Finally, the last phase consists in solving the convexified problem using general-purpose optimization software. 

Our experiments show that \pqcr~is able to solve to global optimality $6$ unsolved instances of the low auto-correlation binary sequence problem and improves lower and/or upper bounds of $27$ of the $45$ instances available at the \texttt{minlplib} website, in comparison to the other available solvers.

The outline of the paper is the following. In Section 2, we define and present our quadratizations of $(P)$. In Section 3, we introduce our family of convex reformulations and we prove how we compute the best parameters. Then, in Section 4, we present our computational results on polynomial instances of degree $4$ coming from the literature and we discuss different possible quadratizations of $(P)$. Section 5 draws a conclusion.


\section{Phase 1: Quadratization of $(P)$}

In this section, we present how we build equivalent quadratic formulations to $(P)$. The basic idea is to reduce the degree of $f$ to $2$. For this, in each monomial of degree 3 or greater, we simply recursively replace each product of two variables by an additional variable. 

More formally, we define the set of indices of the additional variables $J=\{n+1,..,N\}$, where $N$ is the total number of initial and additional variables. We also define the subsets $\cal{E}_i$ for the initial or additional variable $i$ as follows:

\begin{definition}
For all $i\in I\cup J$, we define $\cal{E}_i$ as the set of indices of the variables whose product is equal to $x_i$:
\begin{itemize}
\item  If $i\in I$, i.e. $x_i$ is an initial variable, then we set $\cal{E}_i=\{i\}$
\item If $i\in J$, i.e. $x_i$ is an additional variable, then there exist $(i_1,i_2)\in (I\cup J)^2$ such that $x_i$ replaces $x_{i_1}x_{i_2}$ and we set $\cal{E}_i=\cal{E}_{i_1}\cup\cal{E}_{i_2}$
\end{itemize}
\qed
\end{definition}

Using these sets, we define a \textit{valid} quadratization as a reformulation with $N$ variables where any monomial of degree at least 3 is replaced by the product of two variables.

\begin{definition}
The sets $J=\{n+1,..,N\}$ and $\{\cal{E}_i,~i\in I\cup J\}$ define a valid quadratization with $N$ variables if, for any monomial $p$ of degree greater than or equal to $3$ (i.e. $\vert \cal{M}_p\vert\geq 3$), there exist $(j,k)\in (I\cup J)^2$ such that $\cal{M}_p=\cal{E}_j\cup\cal{E}_k$ and $\prod\limits_{\substack{i\in\cal{M}_p}}x_i=x_jx_k$. Then the monomial $p$ is replaced by a quadratic term.

\qed
\end{definition}

With this definition of a quadratization, we reformulate $(P)$ as a non-convex quadratically constrained quadratic program $(QCQP)$ with $N$ variables. 

\begin{numcases}{(QCQP)}
\min  g(x)=\sum\limits_{\substack{\vert\cal{M}_p\vert\geq 3\\\cal{M}_p=\cal{E}_j\cup\cal{E}_k}}c_px_jx_k +\sum_{\vert\cal{M}_p\vert\leq 2}c_p\prod_{i\in\cal{M}_p}x_i\nonumber\\
\mbox{s.t.}  \nonumber\\
\quad x_i=x_{i_1}x_{i_2} \quad  \forall (i,i_1,i_2) \in J \times (I \cup J)^2: ~\mathcal{E}_i=\mathcal{E}_{i_1}\cup\mathcal{E}_{i_2}\label{fort}\\
 \quad  x \in \{0,1\}^N   \nonumber
 \nonumber
\end{numcases} 

As the variables are binary, Constraints~(\ref{fort}) are equivalent to the classical set of Fortet inequalities~\cite{For59}:

\begin{numcases}{(C_{i_1,i_2}^i)}
 x_i - x_{i_1}\leq 0, &   \nonumber\\
 x_i -x_{i_2}\leq 0 ,&  \nonumber\\
 -x_i + x_{i_1}+x_{i_2}\leq 1, &  \nonumber\\
 -x_i \leq 0, &  \nonumber
\end{numcases} 
We now define set $\mathcal{F}_{\cal{E}}$:
$$\mathcal{F}_{\cal{E}}=\{x\in\{0,1\}^N:  C_{i_1,i_2}^i \textrm{ is satisfied }\forall (i,i_1,i_2) \in J \times (I \cup J)^2: ~\mathcal{E}_i=\mathcal{E}_{i_1}\cup\mathcal{E}_{i_2}\}.$$

We denote by $M=4(N-n)$ the number of constraints of $\mathcal{F}_{\cal{E}}$. We thus obtain the following linearly constrained quadratic formulation that is equivalent to $(P)$ and has $N$ variables and $M$ constraints:

\begin{numcases}{(QP)}
\min  g(x) \equiv x^TQx+c^tx\nonumber\\
\mbox{s.t.}  \nonumber\\
\quad x\in \mathcal{F}_{\cal{E}} \nonumber
\end{numcases} 

\noindent where $Q\in S_N$ (the set of $N\times N$ real symmetric matrices), and $c\in \bb{R}^N$.

\bigskip
In the following, we will focus on the solution of problem $(QP)$ that is an equivalent formulation to $(P)$. Let us observe that $(QP)$, as well as $(QCQP)$, are parameterized by the quadratization defined by sets $\cal{E}$. Indeed, several valid quadratizations can be applied to $(P)$, each of them leading to different sets $\cal{E}_i$. 

\bigskip
Different valid quadratizations were introduced and compared from the size point of view in~\cite{AnBoCrGr17}.  In our case the comparison criterion is the continuous relaxation bound value from which we present our experimental comparison in Section~4.

\bigskip
\begin{example1}[Different valid quadratizations]
~\\Let us consider the following problem: 
$$ (Ex) \Big\{ \min_{x \in \{0,1\}^4} 2x_1+3x_2x_3-2x_2x_3x_4-3x_1x_2x_3x_4$$

\noindent For instance, we can build three different equivalent functions:
  \begin{itemize}
\item $g_1(x) = 2x_1+3x_2x_3-2\underbrace{x_2x_4}_{x_5}x_3-3\underbrace{x_1x_4}_{x_6}\underbrace{x_2x_3}_{x_7}$
\item $g_2(x) =  2x_1+3x_2x_3-2\underbrace{x_3x_4}_{x_5}x_2-3\underbrace{x_1x_2}_{x_6}\underbrace{x_3x_4}_{x_5}$
\item $g_3(x) = 2x_1+3x_2x_3-2\underbrace{x_2x_4}_{x_5}x_3-3\underbrace{x_1x_2}_{x_6}\underbrace{x_3x_4}_{x_7}$
  \end{itemize}
\begin{scriptsize}
  \begin{tabular}{ccc}
\begin{minipage}[c]{.33\linewidth}
\begin{numcases}{(QEx_{1})}
\min g_1(x) \nonumber\\
\mbox{s.t.}\nonumber\\
\quad (x_2,x_4,x_5)\in C_{2,4}^5 \nonumber\\ 
\quad (x_1,x_4,x_6)\in C_{1,4}^6\nonumber\\ 
\quad (x_2,x_3,x_7)\in C_{2,3}^7\nonumber\\ 
\quad  x \in \{0,1\}^{7}  \nonumber
\end{numcases} 
\end{minipage}
&
\begin{minipage}[c]{.33\linewidth}
\begin{numcases}{(QEx_{2})}
\min g_2(x) \nonumber\\
\mbox{s.t.}\nonumber\\
\quad (x_3,x_4,x_5)\in C_{3,4}^5 \nonumber\\ 
\quad (x_1,x_2,x_6)\in C_{1,2}^6\nonumber\\ 
\quad  x \in \{0,1\}^{6}  \nonumber
\end{numcases} 
\end{minipage}

&
\begin{minipage}[c]{.33\linewidth}
\begin{numcases}{(QEx_{3})}
\min g_3(x) \nonumber\\
\mbox{s.t.}\nonumber\\
\quad (x_2,x_4,x_5)\in C_{2,4}^5 \nonumber\\ 
\quad (x_1,x_2,x_6)\in C_{1,2}^6\nonumber\\ 
\quad (x_3,x_4,x_7)\in C_{3,4}^7\nonumber\\ 
\quad  x \in \{0,1\}^{7}  \nonumber
\end{numcases} 
  \end{minipage}
    \end{tabular}
\end{scriptsize}

\bigskip
Here we obtain 3 different quadratizations of $(Ex)$ with different sets $\cal{E}$. They have different sizes: $(QEx_{1})$ and $(QEx_{3})$ have 7 variables and 12 constraints, while $(QEx_{2})$ has 6 variables and 8 constraints.

\end{example1}

We have reduced the degree of the polynomial program $(P)$ by building an equivalent quadratic program to $(P)$. However, the solution of $(QP)$ still has two difficulties, the non-convexity of the objective function $g$ and the integrality of the variables.

\bigskip

Some state-of-the-art solvers can solve $(QP)$ to global optimality (e.g. \texttt{Cplex 12.7}~\cite{cplex127}). Unfortunately, these solvers may not be enough efficient for solving dense instances of $(P)$. Here, we propose to compute an equivalent quadratic convex formulation to $(QP)$. There exist several convexification methods devoted to quadratic programming (see, for example~\cite{BEL16,BEP09,Carter84,HamRub70,McYor80}). These approaches can be directly applied to $(QP)$. For instance, one can use the \texttt{QCR} method, described in~\cite{BEP09}, that consists in computing an equivalent convex formulation to $(QP)$ using semi-definite programming. The convexification is obtained thanks to a non uniform perturbation of the diagonal of the Hessian matrix. The semi-definite relaxation used can be easily solved due to its reasonable size. However, the bound obtained by continuous relaxation of the reformulation is very weak. As a consequence, for the considered instances of Section~4, the \bab~used to solve the reformulation failed as soon as $n\geq 20$. Another alternative is to apply the \texttt{MIQCR} method~\cite{BEL16}. In this method, the perturbation is generalized to the whole Hessian matrix and hence is more refined than the previous one. This leads to a reformulation with a significantly sharper bound. Unfortunately, the semi-definite relaxation used in this approach is too large and its computation failed even with instances of $(P)$ containing only 10 variables. In the next section, we present a new convexification that leads to sharper bounds than $\texttt{QCR}$ but with a better tractability than $\texttt{MIQCR}$.

\section{Phase 2: A quadratic convex reformulation of $(QP)$}

In this section, we consider the problem of reformulating $(QP)$ by an equivalent quadratic 0-1 program with a convex objective function. To do this, we define a new convex function which value is equal to the value of $g(x)$, but which Hessian matrix is positive semi-definite. More precisely, we first add to $g(x)$ a combination of four sets of functions that vanish on the feasible set $\cal{F}_{\cal{E}}$. For each function we introduce a scalar parameter. Then we focus on computing the best parameters that lead to a convex function and that maximize the optimal value of the continuous relaxation of the obtained problem. 

\subsection{Valid quadratic equalities for $(QP)$}

For a quadratization characterized by $\cal{E}$, we introduce null quadratic functions over the set $\mathcal{F}_{\cal{E}}$.

\begin{lemme} The following quadratic equalities characterize null  functions over $\mathcal{F}_{\cal{E}}$:
\begin{small}
\begin{numcases}{(S_\cal{E})}
x_i^2-x_i=0 & $i\in I\cup J$ \label{eq1}\\
x_i-x_ix_j=0& $ (i,j) \in  J \times (I\cup J): ~\mathcal{E}_j\subset\mathcal{E}_i$  \label{eq2}\\
x_i-x_jx_k=0  & $(i,j,k)\in  J \times (I\cup J)^2:  ~\mathcal{E}_i=\mathcal{E}_j\cup\mathcal{E}_k$ \label{eq3} \\
x_ix_j-x_kx_l=0 & $(i,j,k,l)\in  (I\cup J)^4: ~\mathcal{E}_i\cup\mathcal{E}_j=\mathcal{E}_k\cup\mathcal{E}_l$ \label{eq4}
\end{numcases}
\end{small}
\end{lemme}
\begin{preuve} Constraints~(\ref{eq1}) trivially hold since $x_i \in \{0,1\}$. Constraints~(\ref{eq3}) come from Definition~1. We then prove the validity of the Constraints~(\ref{eq2}) and~(\ref{eq4}). 

  \begin{itemize}
    \item \textit{ Constraints~(\ref{eq2}):} we have $x_i=\displaystyle{\prod_{i'\in\cal{E}_i}}x_{i'}$ and $x_j=\displaystyle{\prod_{j'\in\cal{E}_j}x_{j'}}$, then:
\begin{eqnarray}
   x_ix_j & = & \prod_{i'\in\cal{E}_i}x_{i'} \prod_{j'\in\cal{E}_j}x_{j'}\nonumber\\
   & = &  \prod_{j'\in\cal{E}_j}x_{j'}^2\prod_{i'\in\cal{E}_i\backslash\cal{E}_j}x_{i'}  \textrm{ since } \cal{E}_j\subset\cal{E}_i\nonumber\\ 
   & = & \prod_{i'\in\cal{E}_i}x_{i'} \textrm{ since } x^2_{j'} = x_{j'} \textrm{ and } \cal{E}_j \cup (\cal{E}_i\backslash\cal{E}_j) = \cal{E}_i\nonumber\\
   & = & x_i\nonumber
\end{eqnarray}

\item  \textit{Constraints~(\ref{eq4}):} by definition we have:
\begin{eqnarray}
   x_ix_j & = & \prod_{i'\in\cal{E}_i}x'_{i}\prod_{j'\in\cal{E}_j}x_{j'} \nonumber\\
   & = &  \prod_{i'\in\cal{E}_i\cup\cal{E}_j}x_{i'}\prod_{j'\in\cal{E}_i\cap\cal{E}_j}x_{j'} \nonumber\\ 
   & = &  \prod_{i'\in (\cal{E}_i\cup\cal{E}_j) \backslash (\cal{E}_i\cap\cal{E}_j)}x_{i'}\prod_{j'\in\cal{E}_i\cap\cal{E}_j}x^2_{j'} \nonumber\\ 
   &= & \prod_{i'\in\cal{E}_i\cup\cal{E}_j}x_{i'} \textrm{ since } x^2_{j'} = x_{j'} \textrm{ and } (\cal{E}_i\cup \cal{E}_j) \backslash (\cal{E}_i\cap\cal{E}_j) \cup  (\cal{E}_i\cap\cal{E}_j) = (\cal{E}_i\cup\cal{E}_j)  \nonumber\\
   & = & \prod_{k'\in\cal{E}_k\cup\cal{E}_l}x_{k'} \textrm{ since }\cal{E}_i\cup\cal{E}_j = \cal{E}_k\cup\cal{E}_l \nonumber\\
   & = & x_kx_l\nonumber
\end{eqnarray}
\end{itemize}
\end{preuve}

\subsection{An equivalent quadratic convex reformulation to $(QP)$}

We now compute a quadratic convex reformulation of $(QP)$ and thus of $(P)$. For this, we add to the objective function $g$ the null quadratic forms in~(\ref{eq1})--(\ref{eq4}). For each of them, we associate a real scalar parameter:  $\alpha_i$  for Constraints~(\ref{eq1}), $\beta_{ij}$  for Constraints~(\ref{eq2}), $\delta_{ijk}$ for Constraints~(\ref{eq3}), and $\lambda_{ijkl}$ for Constraints~(\ref{eq4}). We get the following parameterized function: 
\begin{eqnarray}
  g_{\alpha,\beta,\delta,\lambda}(x) & = & g(x)+\sum_{i \in I \cup J}\alpha_i(x_i^2-x_i)+\sum\limits_{\substack{(i,j)\in J \times (I \cup J)\\ \cal{E}_j\subset\cal{E}_i}}\beta_{ij}(x_i-x_ix_j) \nonumber \\
  & & +\sum\limits_{\substack{(i,j,k)\in J \times (I\cup J)^2 \\ \mathcal{E}_i=\mathcal{E}_j\cup\mathcal{E}_k}}\delta_{ijk}(x_i-x_jx_k) +\sum\limits_{\substack{(i,j,k,l)\in  (I\cup J)^4 \\ \mathcal{E}_i\cup\mathcal{E}_j=\mathcal{E}_k\cup\mathcal{E}_l}}\lambda_{ijkl}(x_ix_j-x_kx_l) \nonumber 
\end{eqnarray}
Obviously $g_{\alpha,\beta,\delta,\lambda}(x)$ has the same value as $g(x)$ for any $x\in\cal{F_E}$. Moreover, there exist vector parameters $\alpha$, $\beta$, $\delta$ and $\lambda$ such that $g_{\alpha,\beta,\delta,\lambda}$ is a convex function. Take for instance, $\alpha$ equals to the opposite of the smallest eigenvalue of $Q$, and $\beta=\delta=\lambda=0$.

By replacing $g$ by the new function, we obtain the following family of quadratic convex equivalent formulation to $(QP)$:

\begin{numcases}{(QP_{\alpha,\beta,\delta,\lambda})}
\min  g_{\alpha,\beta,\delta,\lambda}(x)\equiv x^TQ_{\alpha,\beta,\delta,\lambda}x+c^T_{\alpha,\beta,\delta,\lambda}x\nonumber\\
\mbox{s.t.}  \nonumber\\
\quad x\in\mathcal{F}_{\cal{E}}\nonumber
 \nonumber
\end{numcases}
\noindent where  $Q_{\alpha,\beta,\delta,\lambda} \in S_N$ is the Hessian matrix of $g_{\alpha,\beta,\delta,\lambda}(x)$, and $c_{\alpha,\beta,\delta,\lambda} \in \mathbb{R}^N$ is the vector of linear coefficients of $g_{\alpha,\beta,\delta,\lambda}(x)$.  


\bigskip

In order to use $(QP_{\alpha,\beta,\delta,\lambda})$ within a \bab~procedure, we are interested by parameters ($\alpha$, $\beta$, $\delta$, $\lambda$) such that $g_{\alpha,\beta,\delta,\lambda}$ is a convex function. Moreover, in order to have a good behavior of the \bab~algorithm, we want to find parameters that give the tightest continuous relaxation bound. More formally, we want to solve the following optimization problem:

$$(CP):\max\limits_{\substack{\alpha\in\mathbb{R}^N,\beta\in\mathbb{R}^{T_1},\delta\in\mathbb{R}^{T_2},\lambda\in\mathbb{R}^{T_3}\\ Q_{\alpha,\beta,\delta,\lambda}\succeq 0}}~~\Big\{\min\limits_{\substack  x\in \overline{\mathcal{F}}_{\cal{E}}} ~~g_{\alpha,\beta,\delta,\lambda}(x) ~\Big\}$$

\noindent where $T_1$, $T_2$ and $T_3$ are the number of Constraints~(\ref{eq2}),~(\ref{eq3}), and~(\ref{eq4}), respectively, and $\overline{\mathcal{F}}_{\cal{E}}$ is the set $\mathcal{F}_{\cal{E}}$ where the integrality constraints are relaxed, i.e. $x \in [0,1]^N$. 
\bigskip

In the rest of the paper we will focus on solving $(CP)$. For this, we build a compact semi-definite relaxation that uses our new valid equalities and prove that its optimal dual variables provide an optimal solution to $(CP)$.

\subsection{Computing an optimal solution to $(CP)$}

The following theorem shows that problem $(CP)$ is equivalent to the dual of a semi-definite relaxation of $(QP)$.

\begin{theorem} The optimal value of $(CP)$ is equal to the optimal value of the following semi-definite program $(SDP)$:
\begin{numcases}{(SDP)}
\min  <Q,X>+c^Tx\nonumber\\
\mbox{s.t.}  \nonumber\\
\quad X_{ii}-x_i=0 & $i\in I\cup J$\label{cont1}\\
 \quad -X_{ij} + x_i=0 &$ (i,j) \in  J \times (I\cup J): ~\mathcal{E}_j\subset\mathcal{E}_i$ \label{cont2}\\
\quad -X_{jk} + x_i=0 &  $(i,j,k)\in  J \times (I\cup J)^2:  ~\mathcal{E}_i=\mathcal{E}_j\cup\mathcal{E}_k$\label{cont3} \\
\quad X_{ij}-X_{kl}=0 & $(i,j,k,l)\in  (I\cup J)^4: ~\mathcal{E}_i\cup\mathcal{E}_j=\mathcal{E}_k\cup\mathcal{E}_l$ \label{cont4}\\
 \quad  \left(
\begin{array}{c c}
1 & x^T\\
x & X
\end{array} \right)\succeq 0 \label{cont5}\\
 \quad x\in\mathbb{R}^N,~X\in S^N\label{cont6}
\end{numcases} 
 
The optimal values $(\alpha^*,\beta^*,\delta^*,\lambda^*)$ of problem $(CP)$ are given by the optimal values of the dual variables associated with constraints~(\ref{cont1})--(\ref{cont4}) respectively.

\end{theorem}

\begin{preuve}
For simplicity, we rewrite $\mathcal{F}_{\cal{E}}$ as follows:  $\mathcal{F}_{\cal{E}}=\{x\in\{0,1\}^N: Ax\leq b\}$ where $A$ is a $M \times N$-matrix, $b \in \mathbb{R}^M$, and we introduce $T=N+T_1+T_2+T_3$ the number of Constraints (\ref{eq1})--(\ref{eq4}) respectively.

\bigskip
We start by observing that $x\in [0,1]^N$ is equivalent to $x^2\leq x$, thus, $(CP)$ is equivalent to $(Q1)$:
$$(Q1):\max\limits_{\substack{\alpha\in\mathbb{R}^N,\beta\in\mathbb{R}^{T_1},\delta\in\mathbb{R}^{T_2},\lambda\in\mathbb{R}^{T_3}\\ Q_{\alpha,\beta,\delta,\lambda}\succeq 0}}~~\Big\{\min\limits_{\substack{x\in\bb{R}^N}, ~x^2 \leq x,~Ax\leq b} ~g_{\alpha,\beta,\delta,\lambda}(x)~\Big\}$$

$(Q1)$ is a convex optimization problem over a convex set. If we consider the solution $\tilde{x}_i=0.5$ $\forall i\in I$ and $\tilde{x}_i=\tilde{x}_j\tilde{x}_k$ $\forall (i,j,k) \in J \times (I \cup J)^2,~\mathcal{E}_i=\mathcal{E}_j\cup\mathcal{E}_k$, the obtained $\tilde{x}$ is an interior point and the Slater's conditions are satisfied for the minimization sub-problem. Then, by Lagrangian duality, we have $(Q1)$ equivalent to $(Q2)$:

$$(Q2):\max\limits_{\substack{\alpha\in\mathbb{R}^N,\beta\in\mathbb{R}^{T_1},\delta\in\mathbb{R}^{T_2},\lambda\in\mathbb{R}^{T_3},\omega\in\bb{R}^N_+, \gamma\in\bb{R}^M_+\\ Q_{\alpha,\beta,\delta,\lambda}\succeq 0}}~~\Big\{\min\limits_{\substack{x\in\bb{R}^N}}~g_{\alpha,\beta,\delta,\lambda}(x)+ \omega^T (x^2 -x)+ \gamma^T (Ax-b)~\Big\}$$

Due to Constraints~(\ref{eq1}), it holds that $(Q2)$ is equivalent to $(Q3)$:

$$(Q3):\max\limits_{\substack{\alpha\in\mathbb{R}^N,\beta\in\mathbb{R}^{T_1},\delta\in\mathbb{R}^{T_2},\lambda\in\mathbb{R}^{T_3}, \gamma\in\bb{R}^M_+\\ Q_{\alpha,\beta,\delta,\lambda}\succeq 0}}~~\Big\{\min\limits_{\substack{x\in\bb{R}^N}} g_{\alpha,\beta,\delta,\lambda}(x) +\gamma^T (Ax-b) ~\Big\}$$

It is well known that a necessary condition for the quadratic function $g_{\alpha,\beta,\delta,\lambda,\gamma}(x)+\gamma^T (Ax-b)$ to have a minimum not equal to $-\infty$ is that matrix $Q_{\alpha,\beta,\delta,\lambda}$ is positive semi-definite. Therefore $(Q3)$ is equivalent to $(Q4)$:

$$(Q4):\max\limits_{\substack{\alpha\in\mathbb{R}^N,\beta\in\mathbb{R}^{T_1},\delta\in\mathbb{R}^{T_2},\lambda\in\mathbb{R}^{T_3}, \gamma\in\bb{R}^M_+}}~~\Big\{\min\limits_{\substack{x\in\bb{R}^N}}g_{\alpha,\beta,\delta,\lambda,\gamma}(x)+\gamma^T (Ax-b)~\Big\}$$

We know from~\cite{LemOus99} that $(Q4)$ is equivalent to problem $(D)$:

\begin{numcases}{(D)}
\max  t\nonumber\\
\mbox{s.t.}  \nonumber\\
 \left(
\begin{array}{c c}
-\gamma^Tb-t & \frac{1}{2}(c^T_{\alpha,\beta,\delta,\lambda}+\gamma^TA)\\
\frac{1}{2}(c_{\alpha,\beta,\delta,\lambda}+A^T\gamma) & Q_{\alpha,\beta,\delta,\lambda}
\end{array} \right) \succeq 0 \nonumber\\
 \quad t\in\mathbb{R},~\alpha\in\mathbb{R}^N,~\beta\in\mathbb{R}^{T_1},~\delta\in\mathbb{R}^{T_2},~\lambda\in\mathbb{R}^{T_3}, ~\gamma\in\bb{R}^M_+\nonumber
 \nonumber
\end{numcases} 

By semi-definite duality of program $(D)$, and with $\alpha,\beta,\delta,\lambda$ the dual variables associated with Constraints~(\ref{cont1})--(\ref{cont4}) respectively, we get $(SDP')$:

\begin{numcases}{(SDP')}
\min  <Q,X>+c^Tx\nonumber\\
\mbox{s.t.}  \nonumber\\
\quad (\ref{cont1})-(\ref{cont6}) \nonumber \\
\quad Ax\leq b\nonumber
\end{numcases} 

\bigskip
We now prove that there is no duality gap between $(D)$ and $(SDP')$, which holds since:
\begin{enumerate}[(i)]
\item The feasible domain of $(SDP')$ is nonempty, as $(QP_{\alpha,\beta,\delta,\lambda})$ contains 0 as a feasible solution and $(D)$ is bounded
\item $(D)$ satisfies Slater's condition. It is sufficient to take $\beta$, $\delta$ and $\lambda$ equal to $0$, $\alpha$ large enough so that $Q_{\alpha,\beta,\delta,\lambda} \succeq 0$ holds, and $t$ a large negative number that ensures the diagonal dominance of the first row and the first column of matrix $\left(
\begin{array}{c c}
-\gamma^Tb-t & \frac{1}{2}(c^T_{\alpha,\beta,\delta,\lambda}+\gamma^TA)\\
\frac{1}{2}(c_{\alpha,\beta,\delta,\lambda}+A^T\gamma) & Q_{\alpha,\beta,\delta,\lambda}
\end{array} \right).$
\end{enumerate}

\bigskip
From these equivalences, we know that we can build an optimal solution of $(CP)$ from the optimal dual variables of $(SDP')$. However, constraints $Ax\leq b$ are redundant in $(SDP')$ and we thus prove in Lemma \ref{theorem_equi} that $(SDP')$ and $(SDP)$ are equivalent. As a consequence, an optimal solution to $(CP)$ can be deduced from the optimal dual variables of $(SDP)$.

\begin{lemme}\label{theorem_equi}
Due to Constraints~(\ref{cont1})--(\ref{cont3}) and (\ref{cont5}), inequalities $Ax \leq b$ are redundant in $(SDP')$.
\end{lemme}
\begin{preuve}
Recall that $Ax \leq b$ are the inequalities of $(C_{j,k}^i), \, \forall (i,j,k) \in J \times (I \cup J)^2: ~\mathcal{E}_i=\mathcal{E}_j\cup\mathcal{E}_k $, i.e $x_i\geq 0$ $(a)$, $x_i\leq x_j$ $(b)$, $x_i\leq x_k$ $(c)$, and $x_i\geq x_j+x_k-1$ $(d)$.

The basic idea used here is that, since matrix $ \left( \begin{array}{c c}1 & x^T\\x & X \end{array} \right)$ is positive semi-definite, all its symmetric minors are non-negative. 
\begin{itemize}
\item Constraint $(a)$: $x_i\geq 0$. We consider the determinant $
\begin{vmatrix}
    1 & x_{i} \\ 
    x_{i} & X_{ii}
  \end{vmatrix}$, which implies $X_{ii}-x_i^2\geq 0$. By (\ref{cont1}) we obtain $x_i-x_i^2\geq 0$ and thus $x_i\geq 0$.
  
\item Constraint $(b)$: $x_i\leq x_j$. Considering the determinant of the symmetric minor 
$\begin{vmatrix}
    X_{jj} & X_{ji} \\ 
    X_{ij} & X_{ii}
\end{vmatrix}$
  implies
  $X_{ii}X_{jj}-X_{ij}^2\geq 0$. By (\ref{cont1}) we have $x_jx_i-X_{ij}^2\geq 0$ and by (\ref{cont2}) we obtain $x_ix_j-x_i^2\geq 0$. Either $x_i> 0$ and then we have $x_j-x_i\geq 0$, or $x_i=0$ and the inequality comes from $x_j\geq 0$.
  
\item Constraint $(c)$: $x_i\leq x_k$. By symmetry, i.e. considering the determinant
  $\begin{vmatrix}
    X_{kk} & X_{ki} \\ 
    X_{ik} & X_{ii}
  \end{vmatrix}$~, the inequality holds.
  
  \item Constraint $(d)$: $x_i\geq x_j+x_k-1$. By definition (\ref{cont5}) implies $z^T \left( \begin{array}{c c} 1 & x^T\\ x & X \end{array} \right)z\geq ~0,\\\forall z\in \bb{R}^{N+1}$. By taking $\bar{z}=(1,0,..,0,\underbrace{-1}_j,0,..,0,\underbrace{-1}_k,0,..,0,\underbrace{1}_i,0,..,0)$, we have:
 
\begin{eqnarray}
 0\leq \bar{z}^T \left(
\begin{array}{c c}
1 & x^T\\
x & X
\end{array} \right)\bar{z} & =& (x_i+1-x_j-x_k)-(x_j-X_{jj}-X_{kj}+X_{ij}) \nonumber\\
& & -(x_k-X_{kk}-X_{jk}+X_{ik})+(x_i-X_{ji}-X_{ki}+X_{ii}) \nonumber\\
&  =&(x_i+1-x_j-x_k) \text{ by (\ref{cont1}), (\ref{cont2}) and (\ref{cont3})} \nonumber.
\end{eqnarray}
\end{itemize}
\end{preuve}

\bigskip
Let us state Corollary~1 that shows that from an optimal dual solution to $(SDP)$ we can build an optimal solution to $(CP)$. 
\begin{corollary}
We have $v(CP)=v(SDP)$ where $v(.)$ is the optimal value of problem $(.)$. Consequently, an optimal solution  $(\alpha^*,\beta^*,\delta^*,\lambda^*)$ of $(CP)$ corresponds to the optimal values of the dual variables associated with constraints~(\ref{cont1})--(\ref{cont4}) of $(SDP)$ respectively.
\end{corollary}

\begin{preuve}
We have:
\begin{enumerate}[(i)]
\item $v(CP)= v(D)$ 
\item since there is no duality gap between $(D)$ and $(SDP')$, we have $v(D)= v(SDP')$
\item by Lemma~2, we get $v(CP)= v(D)=v(SDP')=v(SDP)$
\end{enumerate}

\end{preuve}

\end{preuve}
\bigskip

To sum up, we obtain $(QP^*)$, the best equivalent convex formulation to $(QP)$: 

\begin{numcases}{(QP^*)}
\min  g_{\alpha^*,\beta^*,\delta^*,\lambda^*}(x)\nonumber\\
\mbox{s.t.}  \nonumber\\
\quad x\in\mathcal{F}_{\cal{E}}\nonumber
 \nonumber
\end{numcases}

From Theorem~1, we deduce the Algorithm~1 to solve $(P)$.
\begin{algorithm}
\caption{\pqcr~an exact solution method for $(P)$}
\begin{algorithmic}

\STATE \textbf{Step 1:} Apply a quadratization $\cal{E}$ to $(P)$ and thus generate sets $\mathcal{F}_{\cal{E}}$  and $\mathcal{S}_{\cal{E}}$.\\
\STATE \textbf{Step 2:} Solve $(SDP)$, deduce optimal values $\alpha^*$, $\beta^*$, $\delta^*$, $\lambda^*$, and build $(QP^*)$.\\
\STATE \textbf{Step 3:} Solve $(QP^*)$ by a standard quadratic convex programming solver.

\end{algorithmic}
\end{algorithm}

\section{Numerical results}

In this section, we evaluate \pqcr~on two applications. The first one is the \textit{image restoration (vision)} problem \cite{CraRo17}, which results are presented in Section~4.1. The instances of this application are quite sparse with an average ratio $\frac{m}{n}$ of about $7$. We choose to use these instances in order to compare \pqcr~ with existing convexifications and in particular with methods \texttt{QCR} and \texttt{MIQCR} that are not able to handle larger and/or denser instances. Then, in Section~4.2, we present the results of the second application, the \textit{low auto-correlation binary sequence (LABS)} problem \cite{Berna86} which instances are much denser (average ratio $\frac{m}{n}$ of about $212$). These instances are available on the  \texttt{minlplib} website~\cite{minlplib}, and are very hard to solve. For most of them, the optimal solution value is not known. For these experiments, we have chosen the quadratization described in Algorithm~2 for Step~1 of \pqcr. This choice impacts the number of constraints within the sets $\mathcal{F}_{\cal{E}}$ and $\mathcal{S}_{\cal{E}}$, and the associated continuous relaxation bound value can thus vary. We further illustrate this variation on toy instances in Section~4.3. \\

The quadratization used in our experiments is presented in Algorithm~2.
\begin{center}
  \begin{algorithm}[H]
  \caption{Quadratization($f$)}
\begin{algorithmic}
\REQUIRE A polynomial $f$ of degree $d>2$
\ENSURE A quadratic function $f'$ verifying $\forall x\in\{0,1\}^n, ~f'(x)=f(x)$
\vspace{0.5cm}
\FOR {each monomial $p$ from $1$ to $m$}
\STATE Sort $p$ by lexicographical order
\STATE $deg \leftarrow deg(p)$
\WHILE{$deg>2$}
\STATE $s\leftarrow \lfloor\frac{deg}{2}\rfloor$
\FOR {$l$ from $1$ to $s$}
\STATE  Consider the $l^{th}$ consecutive pair of variables $x_jx_k$
\STATE  Find $x_i$ that represents the product $x_jx_k$
\IF {$x_i$ does not exist} 
\STATE Create an additional variable $x_i$ and $\cal{E}_i\leftarrow\cal{E}_j\cup\cal{E}_k$
\ENDIF
\STATE  Replace $x_jx_k$ by $x_i$
\ENDFOR
\STATE $deg \leftarrow \lceil\frac{deg}{2}\rceil$
\ENDWHILE
\ENDFOR
\end{algorithmic}
\end{algorithm}
\end{center}

\begin{example}
Applying the quadratization of Algorithm~2 to the monomial $x_1x_2x_3x_4x_5$ we obtain the following monomial of degree $3$ at the first iteration:

$$\underbrace{x_1x_2}_{x_6}\underbrace{x_3x_4}_{x_7}x_5$$ 

\noindent we then obtain a quadratic reformulation of the monomial at the second iteration using $3$ additional variables:

$$\underbrace{x_6x_7}_{x_8}x_5$$ 
\qed
\end{example}

\par Our experiments were carried out on a server with $2$ CPU Intel Xeon each of them having $12$ cores and $2$ threads of $2.5$ GHz and $4*16$ GB of RAM using a Linux operating system. For all algorithms, we used the multi-threading version of \texttt{Cplex 12.7} with up to 48 threads. \\

\par In our experiments, we use three classes of solution algorithms:
\begin{enumerate}[i)]
\item The first class includes 3-phase algorithms that consist in a quadratization and a convexification followed by the solution of the equivalent convex problem with the solver \texttt{Cplex 12.7}: these methods are \pqcr~and \texttt{Q+QCR}. For both methods, the quadratization is implemented in \texttt{C}, and we used the solver \texttt{csdp} to solve the semi-definite programs. For denser instances (Section~4.2), we used the solver \texttt{csdp}~\cite{csdp} together with the \texttt{Conic Bundle algorithm}~\cite{cb} to solve the semi-definite program of \pqcr, as described in~\cite{BELW17}. Then, we used the \texttt{ampl}~\cite{AMPL94} interface of the solver \texttt{Cplex 12.7}~\cite{cplex127} to solve the obtained quadratic convex problem with binary variables.
  \item The second class includes a 2-phase algorithm, called \texttt{Q+Cplex}, that consists in a quadratization followed by the direct submission to \texttt{Cplex 12.7}. Here again, the quadratization is implemented in \texttt{C}, and we used the \texttt{ampl} interface of the solver \texttt{Cplex 12.7}.
  \item The third class includes the direct submission to the general mixed-integer non-linear solver \texttt{Baron 17.4.1}~\cite{baron}. Here, we used the \texttt{gams} interface of the solver \texttt{Baron 17.4.1}.
\end{enumerate}
\bigskip

\noindent \textit{Parameters of the solvers}
\begin{itemize}
\item \texttt{Cplex} :   we let the default parameters, except the parameter \texttt{qptolin} that is set to \texttt{0} for methods \pqcr~ and  \texttt{Q+QCR}.
\item \texttt{csdp} : Parameters \texttt{axtol, aytol} of \texttt{Csdp} are set to $10^{-3}$.
\item \texttt{Conic Bundle} : the precision is set to $10^{-3}$. Parameter $p$ (see~\cite{BELW17}) is set to $0.2*|\mathcal{F_{E}}|$.
\end{itemize}

\noindent \textit{Legends of Tables \ref{vision}-\ref{minlplib}}
\begin{itemize}
\item \textit{Name}: Name of the considered instance.
\item \textit{n}: number of variables in the polynomial formulation.
\item \textit{m}: number of monomials.
\item  \textit{BKN}: is the optimal solution value or the best known solution value of the instance. 
\item \textit{N}: number of variables after quadratization.
\item \textit{gap}: is the initial gap, i.e. the gap at the root node of the branch-and-bound,
  $gap = \displaystyle{ \left|\frac {BKN - LB_{i}}{BKN} \right| * 100}$, where $LB_{i}$ is the initial lower bound.
\item \textit{Solution}: best  solution value found within the time limit.
\item \textit{tSdp}: CPU time in seconds for solving semi-definite programs in \pqcr~and \texttt{Q+QCR}. The time limit is set to 2400 seconds for the \textit{vision} problem and 3 hours for the \textit{LABS} problem. If the solver reaches the time limit, \textit{tSdp} is labeled as "-".
\item \textit{tTotal}: total CPU time in seconds of the associated method. The time limit is set to 1 hour for the \textit{vision} problem and 5 hours for the \textit{LABS} problem. If an instance remains unsolved within the time limit, we put the \textit{final gap}$ = \displaystyle \left|\frac {BKN - LB_{f}}{BKN} \right| * 100$, where $LB_{f}$ is the final lower bound.
\item \textit{Nodes}: number of nodes visited by the \bab~algorithm.
\end{itemize}

\subsection{The image restoration problem}

The \textit{vision} instances are inspired from the image restoration problem, which arises in computer vision. The goal is to reconstruct an original sharp base image from a blurred image. An image is a rectangle containing $n=l\times h$ pixels. This rectangle is modeled as a binary matrix of the same dimension. A complete description of these instances can be found in~\cite{CraRo17}. The problem is modeled by the minimization of a degree $4$ polynomial of binary variables where each variable represents a pixel. The coefficients of the monomials are indicative of how likely a configuration is to appear on the sharp base image. The size of the considered instances are $l \times h = 10 \times 10$, $10 \times 15$, and $15\times 15$, or in the polynomial formulation $n=100, 150$ and $225$, with a number of monomials of $m= 668$, $1033$, and $1598$ respectively. In our experiments, 15 instances of each size are considered obtaining a total of $45$ instances. Observe that the $15$ instances of the same size have identical monomials with different coefficients, because they represent different images with the same number of pixels.

 We now focus on the comparison of several convexification methods after quadratization. Indeed, several ways are possible to solve the quadratic non-convex  program $(QP)$. For instance, the standard solver \texttt{Cplex} can directly handle it, or one can apply the \texttt{QCR}~\cite{BEP09} or \texttt{MIQCR}~\cite{BEL16} methods. We compare \pqcr~with these three approaches. We do not report the results for method \texttt{Q+MIQCR} since it was not able to start the computation due to the size of the considered instances. We also give the computational results coming from the direct submission of $(P)$ to the solver \texttt{Baron 17.4.1}. Our observations for these instances are summed up in Table~\ref{vision}, where each line corresponds to one instance, where the $i^{th}$ instance of $l\times h$ pixels is labeled \texttt{v.l.h i}. 
%
%
  
  \begin{table}[H]
\centering
  \begin{scriptsize}
\begin{tabular}{|l|c|c|c||c|c|c||c|c|c||c|c||c|c|} \hline
  \multicolumn{4}{|c||}{Instance} &\multicolumn{3}{|c||}{\pqcr}& \multicolumn{3}{|c||}{\texttt{Q+QCR}} & \multicolumn{2}{|c||}{\texttt{Q+Cplex}}&\multicolumn{2}{|c|}{\texttt{Baron}} \\ \hline
  \textit{Name}&\textit{n}&\textit{m}&\textit{N}&\textit{Gap} &\textit{tSdp}& \textit{tTotal} & \textit{Gap}&\textit{tSdp}& \textit{tTotal}&\textit{Gap}&\textit{tTotal}&\textit{Gap}&\textit{tTotal}\\ \hline
  \texttt{v.10.10 1}&100&668&352&\textbf{0,59}&66&68&396&7&(250 \%)&1113&\textbf{2}&1098&15\\ \hline
\texttt{v.10.10 2}&100&668&352&\textbf{0,28}&64&66&536&8&(343 \%)&1549&\textbf{2}&1529&10\\ \hline
\texttt{v.10.10 3}&100&668&352&\textbf{0,05}&65&67&973&8&(573 \%)&3375&\textbf{1}&3332&6\\ \hline
\texttt{v.10.10 4}&100&668&352&\textbf{0,12}&63&65&957&8&(561 \%)&3377&\textbf{1}&3334&6\\ \hline
\texttt{v.10.10 5}&100&668&352&\textbf{0,13}&65&66&1006&8&(585 \%)&3568&\textbf{1}&3523&5\\ \hline
\texttt{v.10.10 6}&100&668&352&\textbf{0,11}&73&74&359&9&(229 \%)&984&\textbf{2}&972&11\\ \hline
\texttt{v.10.10 7}&100&668&352&\textbf{0,02}&64&65&305&8&(194 \%)&829&\textbf{2}&817&14\\ \hline
\texttt{v.10.10 8}&100&668&352&\textbf{1,37}&64&66&1376&8&(804 \%)&4765&\textbf{1}&4705&7\\ \hline
\texttt{v.10.10 9}&100&668&352&\textbf{3,02}&65&67&1749&8&(1026 \%)&6187&\textbf{1}&6110&4\\ \hline
\texttt{v.10.10 10}&100&668&352&\textbf{3,64}&66&68&1879&8&(1075 \%)&6843&\textbf{1}&6757&4\\ \hline
\texttt{v.10.10 11}&100&668&352&\textbf{0,36}&70&72&489&8&(316 \%)&1388&\textbf{2}&1370&35\\ \hline
\texttt{v.10.10 12}&100&668&352&\textbf{0,20}&70&72&361&9&(232 \%)&997&\textbf{2}&984&23\\ \hline
\texttt{v.10.10 13}&100&668&352&\textbf{0,00}&60&61&709&8&(392 \%)&2654&\textbf{1}&2620&2\\ \hline
\texttt{v.10.10 14}&100&668&352&\textbf{0,00}&60&61&546&8&(297 \%)&2027&\textbf{1}&2001&2\\ \hline
\texttt{v.10.10 15}&100&668&352&\textbf{0,00}&118&119&541&8&(285 \%)&2048&\textbf{1}&2022&1\\ \hline
\texttt{v.10.15 1}&150&1033&542&\textbf{0,31}&290&294&447&24&(351 \%)&1245&\textbf{5}&1234&80\\ \hline
\texttt{v.10.15 2}&150&1033&542&\textbf{0,00}&285&287&367&24&(287 \%)&999&\textbf{5}&990&36\\ \hline
\texttt{v.10.15 3}&150&1033&542&\textbf{0,05}&280&283&1027&27&(772 \%)&3549&\textbf{3}&3520&6\\ \hline
\texttt{v.10.15 4}&150&1033&542&\textbf{0,33}&276&280&845&27&(640 \%)&2840&\textbf{3}&2817&7\\ \hline
\texttt{v.10.15 5}&150&1033&542&\textbf{0,07}&269&271&799&27&(595 \%)&2808&\textbf{3}&2785&4\\ \hline
\texttt{v.10.15 6}&150&1033&542&\textbf{0,55}&297&302&462&25&(366 \%)&1277&\textbf{5}&1266&47\\ \hline
\texttt{v.10.15 7}&150&1033&542&\textbf{0,08}&288&291&360&26&(283 \%)&981&\textbf{5}&972&38\\ \hline
\texttt{v.10.15 8}&150&1033&542&\textbf{0,79}&281&284&1792&26&(1356 \%)&6202&\textbf{3}&6152&19\\ \hline
\texttt{v.10.15 9}&150&1033&542&\textbf{1,80}&283&286&1525&26&(1160 \%)&5209&\textbf{3}&5167&10\\ \hline
\texttt{v.10.15 10}&150&1033&542&\textbf{1,38}&275&279&1510&25&(1124 \%)&5500&\textbf{3}&5456&7\\ \hline
\texttt{v.10.15 11}&150&1033&542&\textbf{0,10}&283&286&391&25&(305 \%)&1102&\textbf{5}&1092&41\\ \hline
\texttt{v.10.15 12}&150&1033&542&\textbf{0,60}&275&279&453&25&(355 \%)&1269&\textbf{5}&1258&125\\ \hline
\texttt{v.10.15 13}&150&1033&542&\textbf{0,00}&254&256&634&27&(469 \%)&2254&\textbf{3}&2236&4\\ \hline
\texttt{v.10.15 14}&150&1033&542&\textbf{0,04}&269&273&731&27&(547 \%)&2590&\textbf{3}&2569&2\\ \hline
\texttt{v.10.15 15}&150&1033&542&\textbf{0,00}&258&259&576&28&(423 \%)&2183&\textbf{2}&2165&2\\ \hline
\texttt{v.15.15 1}&225&1598&827&\textbf{0,12}&1234&1244&365&70&(320 \%)&998&\textbf{9}&993&(95 \%)\\ \hline
\texttt{v.15.15 2}&225&1598&827&\textbf{0,43}&1251&1265&482&83&(421 \%)&1350&\textbf{9}&1343&(138 \%)\\ \hline
\texttt{v.15.15 3}&225&1598&827&\textbf{0,10}&1167&1175&678&65&(582 \%)&2326&\textbf{5}&2313&(83 \%)\\ \hline
\texttt{v.15.15 4}&225&1598&827&\textbf{0,04}&1251&1256&877&65&(753 \%)&2996&\textbf{5}&2980&(127 \%)\\ \hline
\texttt{v.15.15 5}&225&1598&827&\textbf{0,03}&1167&1174&641&67&(546 \%)&2252&\textbf{5}&2240&(76 \%)\\ \hline
\texttt{v.15.15 6}&225&1598&827&\textbf{0,28}&1238&1249&403&64&(353 \%)&1104&\textbf{10}&1098&(107 \%)\\ \hline
\texttt{v.15.15 7}&225&1598&827&\textbf{0,36}&1237&1246&525&67&(463 \%)&1455&\textbf{10}&1447&(144 \%)\\ \hline
\texttt{v.15.15 8}&225&1598&827&\textbf{0,29}&1197&1205&1148&73&(979 \%)&4104&\textbf{5}&4082&(137 \%)\\ \hline
\texttt{v.15.15 9}&225&1598&827&\textbf{0,27}&1170&1176&1542&66&(1315 \%)&5570&\textbf{5}&5541&(171 \%)\\ \hline
\texttt{v.15.15 10}&225&1598&827&\textbf{0,31}&1173&1179&1194&67&(1020 \%)&4380&\textbf{5}&4357&1154\\ \hline
\texttt{v.15.15 11}&225&1598&827&\textbf{0,27}&1224&1230&529&69&(462 \%)&1528&\textbf{8}&1520&(144 \%)\\ \hline
\texttt{v.15.15 12}&225&1598&827&\textbf{0,25}&1225&1235&461&68&(414 \%)&1273&\textbf{12}&1266&(133 \%)\\ \hline
\texttt{v.15.15 13}&225&1598&827&\textbf{0,00}&1124&1128&651&63&(551 \%)&2398&\textbf{4}&2385&1239\\ \hline
\texttt{v.15.15 14}&225&1598&827&\textbf{0,02}&1171&1177&651&63&(553 \%)&2359&\textbf{5}&2346&(79 \%)\\ \hline
\texttt{v.15.15 15}&225&1598&827&\textbf{0,00}&1100&1103&609&65&(513 \%)&2320&\textbf{4}&2308&263\\ \hline

\end{tabular}
  \end{scriptsize}
\caption {Comparison of the maximum time on 4 solution methods for the \textit{vision} instances - time limit one hour}
\label{vision}
\end{table}

We start by comparing the convexification phase of our new algorithm with the original \texttt{QCR} and \texttt{MIQCR} methods. We observe that none of these convexifications are able to handle any considered instances: \texttt{QCR} because of the weakness of its initial gap, and  \texttt{MIQCR} because of the size of the semidefinite problem considered for computing the best reformulation. These experiments confirm the interest of designing \pqcr,~an algorithm devoted to polynomial optimisation.
Then, we can see that \texttt{Q+Cplex} dominates \pqcr. However, one have to note that these instances are very sparse (average ratio $\frac{m}{n}$ of about $7$). It is well known that the standard linearization performs very well on sparse instances. Clearly, for these instances, the time spent on solving a large semidefinite program, even once, is not profitable in comparison to the efficiency of LP heuristic or cut methods implemented in \texttt{cplex 12.7}. Indeed,  \texttt{Q+Cplex} solves all the considered instances at the root node of its \bab. Moreover, it is interesting to remark that $99\%$ of the CPU time of \pqcr~is spent for solving $(SDP)$, while the CPU time for solving $(QP^*)$ is always smaller than $14$ seconds.
Finally, we compare \pqcr~with the direct submission to the solver \texttt{Baron}. We observe that \texttt{Baron} is faster than \pqcr~on the medium size instances ($n=100$ or $150$), but is not able to solve all the larger instances within the time limit. Indeed, for $n=225$, it solves only $3$ instances out of $15$. On the contrary, \pqcr~seems quite stable to the increase of the size of the instances. Indeed, the initial gap remains stable ($0.42\%$ on average) while the total CPU time increases reasonably.

\subsection{The \textit{Low Auto-correlation Binary Sequence} problem}
We consider the problem of binary sequences with low off-peak auto-correlations. More formally, let $S$ be a sequence $S = (s_1, \ldots ,s_n)$ with $s \in \{-1, 1\}^n$, and for a given $k=0,\ldots ,n-1$, we define the auto-correlations $C_k(S)$ of $S$: $$C_k(S)=\sum_{i=1}^{n-k}s_is_{i+k}$$
The problem is to find a sequence $S$ of length $n$ that minimizes $E(S)$, a degree $4$ polynomial:
$$E(S)=\sum_{k=1}^{n-1}C^2_k(S)$$

This problem has numerous practical applications in communication engineering, or theoretical physics~\cite{Berna86}. For our experiments, we consider truncated instances, i.e. sequences of length $n$ where we compute low off-peak auto-correlation up to a certain distance $n_0\leq n$, i.e. we consider the following function to minimize:
$$E_{n_0}(S)=\sum_{k=1}^{n_0-1}C^2_k(S)$$
In order to apply \pqcr, which is initially tailored for $\{0,1\}$ polynomial programs, we convert the variables from $\{-1,1\}$ to $\{0,1\}$ using the standard transformation $x=\frac{s+1}{2}$.

This problem admits a lot of symmetries. In particular the correlations $C_k$ are identical for a sequence $S$ and its complement. We exploited this symmetry by fixing to 0 the variable that appears the most. Each instance is labeled \texttt{b.n.n}$_\texttt{0}$.
These instances were introduced by~\cite{LMPRS10} and can be found on the \texttt{minlplib}~\cite{minlplib} or the \texttt{polip}~\cite{polip} websites. We do not report the results for methods \texttt{Q+QCR} and \texttt{Q+MIQCR} since they have failed to solve all the considered instances. Two instances that are already quadratic (\texttt{b.20.03} and \texttt{b.25.03}) are solved by the method \texttt{Q+Cplex} in $7$ and $75$ seconds respectively. However, this method was not able to solve the other instances within the time limit.

\begin{table}[H]
\centering
\begin{tabular}{|l|c|c||c|c|c|c|c||c|c|c|} \hline
  \multicolumn{3}{|c||}{Instance} &\multicolumn{5}{|c||}{\pqcr}& \multicolumn{3}{|c|}{\texttt{Baron 17.4.1}}\\ \hline
\textit{Name}&\textit{n}&\textit{m}&\textit{N}&\textit{Gap} &\textit{tSdp}& \textit{tTotal} &\textit{Nodes}&\textit{Gap}&\textit{tTotal}&\textit{Nodes}\\ \hline
\texttt{b.20.03}&20&38&20&\textbf{0}&1&2&0&100&\textbf{1}&1\\ \hline
\texttt{b.20.05}&20&207&65&\textbf{23}&22&23&5886&1838&\textbf{2}&1\\ \hline
\texttt{b.20.10}&20&833&124&\textbf{8}&837&846&24183&2918&\textbf{125}&7\\ \hline
\texttt{b.20.15}&20&1494&164&\textbf{5}&1228&1242&9130&3202&\textbf{728}&9\\ \hline
\texttt{b.25.03}&25&48&25&\textbf{0}&1&2&0&100&\textbf{0}&1\\ \hline
\texttt{b.25.06}&25&407&105&\textbf{17}&461&469&163903&2307&\textbf{65}&27\\ \hline
\texttt{b.25.13}&25&1782&206&\textbf{4}&\textbf{1552}&1603&76828&3109&3750&75\\ \hline
\texttt{b.25.19}&25&3040&265&\textbf{4}&-&\textbf{13433}&224550&3356&14399&129\\ \hline
\texttt{b.25.25}&25&3677&289&\textbf{5}&-&\textbf{13395}&167423&3405&(12 \%)&100\\ \hline
\texttt{b.30.04}&30&223&82&\textbf{23}&58&78&134635&1347&\textbf{7}&7\\ \hline
\texttt{b.30.08}&30&926&174&\textbf{10}&1940&\textbf{2040}&752765&2696&2778&237\\ \hline
\texttt{b.30.15}&30&2944&296&\textbf{5}&-&\textbf{13525}&438278&3221&(21 \%)&103\\ \hline
\texttt{b.30.23}&30&5376&390&\textbf{11}&5953&\textbf{6865}&9337391&3450&(135 \%)&8\\ \hline
\texttt{b.30.30}&30&6412&422&\textbf{4}&8500&\textbf{15352}&452460&3470&(161 \%)&5\\ \hline
\texttt{b.35.04}&35&263&97&\textbf{19}&135&167&156085&1350&\textbf{32}&13\\ \hline
\texttt{b.35.09}&35&1381&234&\textbf{10}&2245&\textbf{4630}&8163651&2826&(29 \%)&354\\ \hline
\texttt{b.35.18}&35&5002&419&\textbf{644}&-&\textbf{(12 \%)}&4899872&3356&(133 \%)&10\\ \hline
\texttt{b.35.26}&35&8347&530&\textbf{30}&-&\textbf{(5 \%)}&5006407&3508&(229 \%)&3\\ \hline
\texttt{b.35.35}&35&10252&579&\textbf{12}&-&\textbf{(11 \%)}&134426&3499&(214 \%)&3\\ \hline
\texttt{b.40.05}&40&447&145&\textbf{25}&430&\textbf{1630}&23459121&1856&3674&1021\\ \hline
\texttt{b.40.10}&40&2053&304&\textbf{9}&-&\textbf{(4 \%)}&25480163&2953&(54 \%)&147\\ \hline
\texttt{b.40.20}&40&7243&544&\textbf{9}&-&\textbf{(4 \%)}&9783350&3405&(203 \%)&3\\ \hline
\texttt{b.40.30}&40&12690&702&\textbf{360}&-&\textbf{(25 \%)}&281134&3561&(274 \%)&1\\ \hline
\texttt{b.40.40}&40&15384&762&\textbf{62}&-&\textbf{(44 \%)}&57534&3536&(464 \%)&1\\ \hline
\texttt{b.45.05}&45&507&165&\textbf{24}&1384&\textbf{(4 \%)}&84159279&1854&\textbf{16609}&4727\\ \hline
\texttt{b.45.11}&45&2813&382&\textbf{9}&-&\textbf{(2 \%)}&25114985&3018&(132 \%)&33\\ \hline
\texttt{b.45.23}&45&10776&706&\textbf{21}&-&\textbf{(16 \%)}&1225234&3470&(242 \%)&2\\ \hline
\texttt{b.45.34}&45&18348&898&\textbf{137}&-&\textbf{(105 \%)}&38513&3604&(375 \%)&1\\ \hline
\texttt{b.45.45}&45&21993&969&\textbf{187}&-&\textbf{(153 \%)}&25964&3559&(624 \%)&1\\ \hline
\texttt{b.50.06}&50&882&230&\textbf{19}&1230&\textbf{(9 \%)}&49490829&2321&(35 \%)&1225\\ \hline
\texttt{b.50.13}&50&4457&506&\textbf{8}&-&\textbf{(5 \%)}&12039566&3131&(192 \%)&7\\ \hline
\texttt{b.50.25}&50&14412&866&\textbf{676}&-&\textbf{(247 \%)}&684010&3511&(280 \%)&1\\ \hline
\texttt{b.50.38}&50&25446&1118&\textbf{242}&-&\textbf{(163 \%)}&309289&3646&(505 \%)&1\\ \hline
\texttt{b.50.50}&50&30271&1202&\textbf{360}&-&\textbf{(305 \%)}&49507&3541&(729 \%)&1\\ \hline
\texttt{b.55.06}&55&977&255&\textbf{21}&-&\textbf{(11 \%)}&23603952&2323&(54 \%)&6\\ \hline
\texttt{b.55.14}&55&5790&607&\textbf{11}&-&\textbf{(7 \%)}&7829649&3186&(373 \%)&6\\ \hline
\texttt{b.55.28}&55&19897&1069&\textbf{174}&-&\textbf{(106 \%)}&580827&3553&(646 \%)&2\\ \hline
\texttt{b.55.41}&55&33318&1347&\textbf{330}&-&\textbf{(244 \%)}&117912&3654&(639 \%)&1\\ \hline
\texttt{b.55.55}&55&40402&1459&\textbf{547}&-&\textbf{(493 \%)}&117027&3575&(705 \%)&1\\ \hline
\texttt{b.60.08}&60&2036&384&\textbf{12}&-&\textbf{(9 \%)}&24800852&2712&(175 \%)&1\\ \hline
\texttt{b.60.15}&60&7294&716&\textbf{16}&-&\textbf{(14 \%)}&4044387&3236&(404 \%)&1\\ \hline
\texttt{b.60.30}&60&25230&1264&\textbf{256}&-&\textbf{(165 \%)}&295197&3578&(471 \%)&1\\ \hline
\texttt{b.60.45}&60&43689&1614&\textbf{547}&-&\textbf{(439 \%)}&26955&704&(671 \%)&1\\ \hline
\texttt{b.60.60}&60&52575&1742&\textbf{784}&-&\textbf{(704 \%)}&23716&3604&(762 \%)&1\\ \hline
\end{tabular}
\caption {Results of \pqcr~and \texttt{Baron}~for the 45 instances of the LABS problem. Time limit 5 hours. \texttt{-} means that the time limit of 3h on the SDP phase is reached.}
\label{baron}
\end{table}
 We present in Table~\ref{baron} a detailed comparison of \pqcr~with the direct submission to \texttt{baron 17.1.4}. For these experiences the total time limit was set to 5 hours, and we limit the CPU time for solving $(SDP)$ to 3 hours. Indeed, any feasible solution to the dual of $(SDP)$ can be used to get a convex objective function in the equivalent formulation. Thus, if the CPU time in column $tSdp$ is smaller than three hours it means that $(SDP)$ was solved to optimality. In the other case, we get a feasible dual solution and we can suppose that the initial gap of \pqcr~could be improved. For these instances \pqcr~is faster than baron since it solves $17$ instances out of $45$ within the time limit while \texttt{baron} solves only $13$ instances. Here, \texttt{baron 17.1.4} solves $2$ instances that were stated as unsolved on \texttt{minlplib}. As expected, \pqcr~has an initial gap much smaller that \texttt{baron} (reduced by a factor $22$ on average). We also observe that the number of nodes visited by \pqcr~during the \bab~is significantly larger than the the number of nodes of \texttt{baron} (increased by a factor of about $40 000$ on average).

\begin{table}[h]
  \centering
\begin{tabular}{|l||c|c||c|c|} \hline
  \multicolumn{1}{|c||}{Instance} &\multicolumn{2}{|c||}{\pqcr~(5h)}& \multicolumn{2}{|c|}{\texttt{minlplib} \cite{minlplib}}\\ \hline
\textit{Name} &\textit{Solution}& \textit{$LB_f$} &\textit{Solution}&\textit{$LB_f$} \\ \hline
\texttt{b.25.19$^{**}$}&-14644&-14644&-14644&-16108\\ \hline
\texttt{b.25.25$^{**}$}&-10664&-10664&-10664&-12494\\ \hline
\texttt{b.30.15$^{**}$}&-15744&-15744&-15744&-19780\\ \hline
\texttt{b.30.23$^{**}$}&-30460&-30460&-30420&-72030\\ \hline
\texttt{b.30.30$^{**}$}&-22888&-22888&-22888&-54014\\ \hline
\texttt{b.35.09$^{**}$}&-5108&-5108&-5108&-6312\\ \hline
\texttt{b.35.18$^*$}&-31144&-34964&-31160&-74586\\ \hline
\texttt{b.35.26$^\# $$^*$}&-55288&-57789&-55184&-191466\\ \hline
\texttt{b.35.35$^*$}&-41052&-45787&-41068&-290424\\ \hline
\texttt{b.40.10$^\#$$^*$}&-8248&-8551&-8240&-14618\\ \hline
\texttt{b.40.20$^\#$$^*$}&-50576&-52465&-50516&-162365\\ \hline
\texttt{b.40.30$^\#$$^*$}&-94872&-118324&-94768&-398617\\ \hline
\texttt{b.40.40$^*$}&-67528&-98031&-67964&-302028\\ \hline
\texttt{b.45.05$^*$}&-1068&-1112&-1068&-1145\\ \hline
\texttt{b.45.11$^\#$$^*$}&-12748&-13035&-12740&-30771\\ \hline
\texttt{b.45.23$^\#$$^*$}&-85423&-98984&-85248&-320397\\ \hline
\texttt{b.45.34$^*$}&-151352&-311627&-152368&-752427\\ \hline
\texttt{b.45.45$^*$}&-111292&-285811&-112764&-685911\\ \hline
\texttt{b.50.06$^*$}&-2160&-2363&-2160&-2921\\ \hline
\texttt{b.50.13$^\#$$^*$}&-23791&-24975&-23772&-74768\\ \hline
\texttt{b.50.25$^*$}&-124572&-433247&-124748&-562446\\ \hline
\texttt{b.50.38$^*$}&-232344&-611906&-232496&-1318325\\ \hline
\texttt{b.50.50$^*$}&-162640&-681105&-168216&-1173058\\ \hline
\texttt{b.55.06$^*$}&-2400&-2659&-2400&-3439\\ \hline
\texttt{b.55.14$^\#$$^*$}&-33272&-35698&-33168&-116748\\ \hline
\texttt{b.55.28$^*$}&-189896&-392929&-190472&-989145\\ \hline
\texttt{b.55.41$^*$}&-335388&-1160180&-337388&-2494477\\ \hline
\texttt{b.55.55$^*$}&-233648&-1434663&-241912&-1947633\\ \hline
\texttt{b.60.08$^*$}&-6792&-7388&-6792&-13915\\ \hline
\texttt{b.60.15$^\#$$^*$}&-45232&-51467&-44896&-169767\\ \hline
\texttt{b.60.30$^*$}&-259271&-692721&-261048&-1491016\\ \hline
\texttt{b.60.45$^*$}&-475504&-2579935&-478528&-3687344\\ \hline
\texttt{b.60.60$^*$$$}&-343400&-2816441&-350312&-3021077\\ \hline
 \end{tabular}
\caption {Comparison of the best known solution and best lower bound values of \pqcr~and of the \texttt{minlplib}~for the unsolved LABS instances. \texttt{$**$}: solved for the first time, \texttt{$\#$}: best known solution improved, and \texttt{$*$}: best known lower bound improved}
\label{minlplib}
\end{table}

We present in Table~\ref{minlplib} the values of the best solutions and of the final lower bounds obtained by \pqcr~within 5 hours of CPU time, and those available on the \texttt{minlplib} website. More precisely, we report in the column \texttt{minlplib} the best solution/final lower bound value obtained among the results of the solvers \texttt{Antigone}, \texttt{Baron}, \texttt{Couenne}, \texttt{Lindo}, and \texttt{Scip}. \pqcr~solves to optimality 6 unsolved instances (labeled as~$^{**}$). It also improves the best known solution values of 9 instances (labeled as~$^\#$), and improves the final lower bound of all the unsolved instances (labeled as~$^*$). In this table, each line corresponds to one instance, and we only present results for instances that were stated as unsolved on \texttt{minlplib}.
\begin{figure}
\begin{center}

\rotatebox{90}{\includegraphics[scale=0.6]{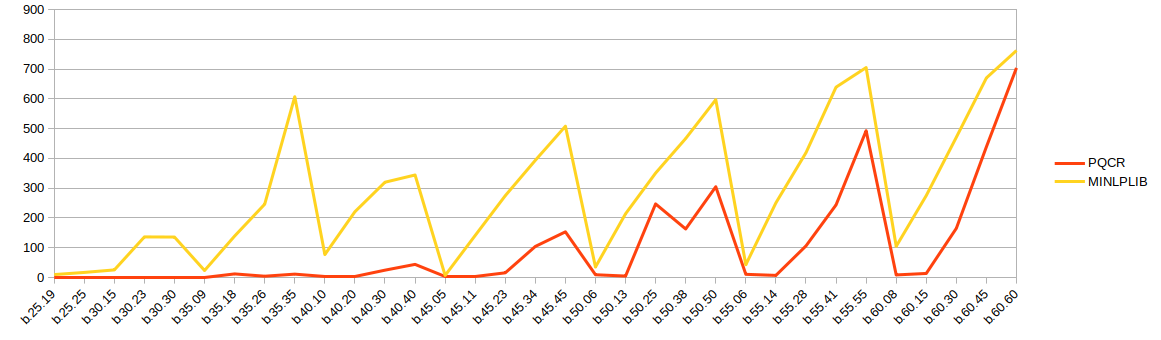}}

\end{center}
  \caption {Comparison between the final gap of \pqcr~and the final gap computed with the best known solution and the best bound from \texttt{minlplib} for the unsolved auto-correlation instances}
  \label{fin_gap}
\end{figure}
 To illustrate these results, we plot in Figure~\ref{fin_gap}, for each instance reported in Table~\ref{minlplib}, the \textit{final gap} of \pqcr~and \texttt{minlplib}. Clearly, the final gap of \pqcr~is much smaller than the final gap of \texttt{minlplib} (reduced by a factor $3$ on average).
 
 \bigskip
 
 A last remark concerns the CPU time necessary to solve $(SDP)$. Indeed, this time represents on average 75\% of the total CPU time. A natural improvement is to identify the set of "important equalities" in a preprocessing step in order to improve the behavior of the solution of $(SDP)$. Obviously, this step should be dependent on the quadratization.

\subsection{A short discussion on the impact of the chosen quadratization}
%
%
%
%
%
%

In this section, we shortly explore the impact of the chosen quadratization on the tightness of the associated continuous relaxation bound. In Table~4, we report the continuous relaxation bound values obtained by convexification after applying the quadratization of Algorithm~2, and three quadratizations from \cite{CraRo17}, namely Pairwise Cover 1, 2 and 3 (PC1, PC2 and PC3). In Pairwise Cover 1, for each monomial of degree $d\geq 3$, the first two variables are linearized to obtain a monomial of degree $d-1$. The process is recursively reproduced until $d=2$. Pairwise Covers 2 and 3 try to minimize the number of additional variables. In PC2, the authors compute the sub-monomials of any degree that appear the most among all the intersection of pairs of monomials. Then they linearize these sub-monomials using the set $\cal{F_E}$ and they repeat the process until the objective function is quadratic. PC3 linearizes in priority the pair of variables that occurs the most frequently in all the monomials. For instance, if we consider the quadratization of the following monomial of degree $4$, $x_1x_2x_3x_4$, we will compute the most frequent pair of variables among the six possible products. If $x_1x_2$ is the most frequent, then the monomial will be quadratized using two variables, one for the reformulation of $x_1x_2$ and the other for $x_3x_4$.

\begin{table}
  \begin{center}
\begin{tabular}{l|c|c|c|c|c|c|c|c|c|}

   &  & \multicolumn{2}{|l|}{$PC1$} & \multicolumn{2}{|l|}{$PC2$} & \multicolumn{2}{|l|}{$PC3$} & \multicolumn{2}{|l|}{$Quad$}\\ 
   \cline{2-10}
  & \textit{Opt} &\textit{N}&\textit{$LB_i$} & \textit{N}&\textit{$LB_i$} &\textit{N}&\textit{$LB_i$} &\textit{N}&\textit{$LB_i$}  \\ \hline
\texttt{b.20.05} & -416 & 64 & -435  & 56&-439&40&-436&65&-435 \\ \hline
\texttt{b.20.10}& -2936 & 123 & -3052  & 135&-3115&93&-3068 &124&-3051 \\ \hline
\texttt{b.40.10} & -8248 & 303 & -8590  & 315&-8659&262&-8745 &304&-8589 \\ \hline
\end{tabular}
\caption{Comparison of bounds and number of variables of \texttt{PQCR} after different quadratizations}
  \end{center}
  \end{table}

We observe that the chosen quadratization impacts $N$, the number of variables of $(QP)$. It also impacts the quality of the associated semidefinite bound, \textit{$LB_i$}. Indeed, the more variables are added, the more the size of sets $\mathcal{F_E}$ and $\mathcal{S_E}$ increases. Clearly, some equalities of $\mathcal{S_E}$ may be stronger than others. Interesting future research directions would be to identify, for a given quadratization, a set of "important" equalities in $\cal{S_E}$, and to determine which quadratization used in \pqcr~leads to faster solution time and/or sharper initial lower bound.

\section{Conclusion}

We consider the general problem $(P)$ of minimizing a polynomial function where the variables are binary. In this paper, we present \pqcr~a solution approach for $(P)$. \pqcr~can be split in 3 phases. We called the first phase \textit{quadratization}, where we rewrite $(P)$ as an equivalent quadratic program $(QP)$. For this we have to add new variables and linear constraints. We get a linearly constrained quadratic program that still has a non-convex objective function and binary variables. Moreover, even for small instances of $(P)$, the existing convexification methods failed to solve the associate $(QP)$.  This is why, we present a family of tailored quadratic convex reformulations of $(QP)$ that exploits its specific structure. For this, we introduce new valid quadratic equalities that vanish on the feasible domain of $(QP)$. We use these equalities to build a family of equivalent quadratic convex formulations to $(QP)$. Then, we focus on finding, within this family, the equivalent convex formulation that maximizes the continuous relaxation bound value. We show that we can compute this "best" convex reformulation using a new semidefinite relaxation of $(QP)$. Finally, we solve our optimal reformulation with a standard solver.

We present computational results on two applications and compare our algorithm with other convexification methods and the general solver \texttt{Baron}. In particular, we show that for the \textit{low auto-correlation binary sequence problem}, \pqcr~is able to improve the best known solution of 10 instances out of 45. A future research direction would be to characterize which quadratization best fit with our convexification phase from the continuous relaxation value point of view.

\bigskip

\subsection*{Acknowledgment}

The authors are thankful to Elisabeth Rodriguez-Heck and Yves Crama for sharing the executable of their quadratization code in order to compute the Pairwise Covers 1, 2 and 3 on \texttt{LABS} instances.

\clearpage

\bibliography{mybib}{}
\bibliographystyle{plain}

\end{document}